\documentclass[twocolumn,epjc3]{svjour3}
\smartqed  
\RequirePackage{graphicx}

\usepackage{amsmath}
\usepackage{xcolor}
\journalname{Eur. Phys. J. C}

\begin{document}

\title{Broad angular anisotropy of multiple scattering in a Si crystal}%

\author{A. Mazzolari \thanksref{addr1} \and A. Sytov \thanksref{e1,addr1,addr2} \and L. Bandiera \thanksref{addr1} \and G. Germogli \thanksref{addr1,addr2} \and M. Romagnoni \thanksref{addr1,addr2} \and E. Bagli \thanksref{addr1} \and V. Guidi  \thanksref{addr1,addr2} \and V. V. Tikhomirov \thanksref{addr3} \and D. De Salvador \thanksref{addr4,addr5} \and S. Carturan \thanksref{addr4,addr5} \and C. Durigello \thanksref{addr4,addr5} \and G. Maggioni \thanksref{addr4,addr5} \and M. Campostrini \thanksref{addr5} \and A. Berra \thanksref{addr6,addr7} \and V. Mascagna \thanksref{addr6,addr7} \and M. Prest \thanksref{addr6,addr7} \and E. Vallazza \thanksref{addr7} \and W. Lauth \thanksref{addr8} \and P. Klag  \thanksref{addr8} \and M. Tamisari  \thanksref{addr1,addr9} }

\institute{INFN Sezione di Ferrara, Via Saragat 1, 44124 Ferrara, Italy \label{addr1}
          \and
          Dipartimento di Fisica e Scienze della Terra, Universit\`{a} di Ferrara, Via Saragat 1, 44124 Ferrara, Italy \label{addr2}
          \and
          Institute for Nuclear Problems, Belarusian State University, Bobruiskaya 11, Minsk 220030, Belarus \label{addr3}
          \and
          Dipartimento di Fisica, Universit\`{a} di Padova, Via Marzolo 8, 35131 Padova, Italy \label{addr4}
          \and          
          INFN Laboratori Nazionali di Legnaro, Viale dell'Universit\`{a} 2, 35020 Legnaro, Italy  \label{addr5}
          \and     
          INFN Sezione di Milano Bicocca, Piazza della Scienza 3, 20126 Milano, Italy  \label{addr6}
          \and               
          Universit\`{a} dell'Insubria, via Valleggio 11, 22100 Como, Italy  \label{addr7}
          \and              
          Institut f\"{u}r Kernphysik der Universit\"{a}t Mainz, 55099 Mainz, Germany  \label{addr8}
          \and              
          Dipartimento di Scienze Biomediche e Chirugico specialistiche, Universit\`{a} di Ferrara, Via Luigi Borsari 46, 44121 Ferrara, Italy  \label{addr9}
}

\thankstext{e1}{e-mail: sytov@fe.infn.it}

\date{Received: date / Accepted: date}

\maketitle

\begin{abstract}
We observed reduction of multiple Coulomb scattering of 855 MeV electrons within a Si crystalline plate w.r.t. an amorphous plate with the same mass thickness. The reduction owed to complete or partial suppression of the coherent part of multiple scattering in a crystal vs crystal orientation with the beam. Experimental data were collected at Mainz Mikrotron and critically compared to theoretical predictions and Monte Carlo simulations. Our results highlighted maximal 7 \% reduction of the r.m.s. scattering angle at certain beam alignment with the [100] crystal axes. However, partial reduction was recorded over a wide range of alignment of the electron beam with the crystal up to 15 deg. This evidence may be relevant to refine the modelling of multiple scattering in crystals for currently used software, which is interesting for detectors in nuclear, medical, high energy physics.

\end{abstract}


\section{Introduction}

Interaction of a charged particle traversing an amorphous medium occurs through repeated interactions with atoms, i.e. the so-called multiple Coulomb scattering \cite{PDG,Mol,Bethe}. This effect is essential for describing any physical experiment connected with the passage of charged particles through matter.

A large variety of detectors, in particular semiconductor detectors, electromagnetic calorimeters, etc. require simulations of multiple scattering using special well-verified software like GEANT 4 \cite{GEANT}, FLUKA \cite{FLUKA} and other codes for event reconstruction. Although the materials which the detectors are frequently made of have a crystalline structure, they are usually treated with a multiple scattering model for an amorphous media. Since a crystal is a solid material whose constituents, such as atoms, molecules or ions, are arranged in a periodical structure forming a lattice, the interaction of particles with a crystal may be different w.r.t. an amorphous medium.

Indeed, particle interaction with atoms of a crystal may occur through two routes: either with individual atoms as for an amorphous medium (incoherent scattering) or with an ordered ensemble of atoms in a crystal as a whole such as atomic strings or planes (coherent scattering). In the latter case, the interaction potentials of individual atoms have to be replaced with the collective potential of atomic strings or planes. Coherent interaction is observed when some conditions for alignment of the particle momentum with the crystal are met, while incoherent interaction is inherently isotropic.

Indeed this is well known in the Diffraction Theory. Diffraction occurs when the particle energy is sufficiently low, and therefore the De Broglie wavelength, $\lambda$, is comparable with the lattice spacing $d_{at}$. Diffraction is usually interpreted in terms of interference patterns in the elastic scattering cross section (coherent interaction), describing the interaction between radiation and the crystal lattice as a whole, i.e. scattering on different atoms cannot be considered independently \cite{Landau}. For instance, diffraction of relatively low energy (10 eV--100 keV) electrons \cite{James} interacting with crystalline solids proved to be a powerful tool to investigate the crystalline structure of matter.

At higher energies, $\lambda$ becomes much shorter than $d_{at}$ and one expects that coherent effects disappear. Nevertheless, at sufficiently high-energy, even if $\lambda \ll d_{at}$, the  typical length of scattering region increases with the energy, being longer than $d_{at}$. Indeed, the length of the scattering region increases with the particle energy, $E$ , embracing more and more atoms along a lattice direction. As for diffraction, the scattering by different atoms is also not independent in this case, and one has to consider the scattering of the particle by the crystal as a whole (coherent scattering). Naturally, coherent interaction adds up with incoherent scattering by individual atoms. 

The first who understood the importance of the lattice structure in the interaction of high-energy particles with crystals were Ferretti and Ter-Mikaelyan, who exploited the Laue theory of diffraction to describe the interference effect of bremsstrahlung in crystals, so-called \textit{coherent bremsstrahlung} (CB). CB is worldwide exploited at accelerator facilities \cite{Ferretti,TM,Uberall}, typically used in hadronic physics \cite{Lohmann,Paterson}. This phenomenon occurs even when the electron and positron incidence angles w.r.t. crystal planes and axes are small (less than $1^\circ$). At very small incidence angle, i.e. lower than the critical one introduced by Lindhard \cite{Lindhard}, a charged particle can be captured in the axial/planar potential well, i.e. \textit{Channeling} occurs \cite{Stark,Davies,Robinson}. For particles moving in a crystal under channeling regime, the multiple scattering encounted is significantly different than if the particle would traverse an amorphous medium with the same density.

As described above, the scattering between the charged particle and crystal can be divided into coherent and incoherent parts. One may expect that in the case of absence of coherent scattering, the total multiple scattering is decreased with respect to an amorphous matter. 

In this paper we introduce the effect of complete coherent scattering suppression (CSS) at certain crystal orientations, which we calculate under proper approximation. We report direct observation of CSS effect, determine the necessary conditions for its manifestation and model it via Monte Carlo simulations. Moreover, we provide an evidence of partial CSS occuring at much broader incidence angle than the critical angle for channeling.

\section{Basics of multiple scattering in a crystal}

As for any situation when coherent and incoherent effects compete, the well known treatment borrowed from X-ray diffraction can be called forward and adapted to our circumstances. We adopted the work that Ter-Mikaelyan carried on for the case of bremsstrahlung \cite{TM} to the case of multiple scattering of charged particles in crystals. We introduced a theoretical calculation to explain CSS following a typical approach for X-ray diffraction \cite{Landau}, being applicable also for classical particle scattering by atomic strings and planes \cite{Baier,Akh}. The particle-crystal interaction potential is described as the sum of individual particle-atom interaction
potentials $U(\mathbf{r})=\sum_j V(|\mathbf{r}-\mathbf{r_j}|)$,
$\mathbf{r}$ being the particle position and $\mathbf{r_j}$ j-th atom
location in the lattice. The latter differs from the equilibrium
atomic position $\mathbf{r_{j0}}$ due to thermal vibrations,
$\mathbf{r_j}=\mathbf{r_{j0}}+\mathbf{u_j}$, $\mathbf{u_j}$ being
the uncorrelated displacement from the lattice nodes,
characterized by the r.m.s. amplitude $<\mathbf{u_j^2}>$. 

As a high-energy particle impinges on a crystal axis or plane at
large enough angle, its wave function resembles a plane
wave. Therefore, the Born approximation is traditionally applied to treat
its interaction with the crystal by using the differential scattering
cross-section ($\hbar=c=1$):

\begin{figure}
 \includegraphics[width=0.35\textwidth]{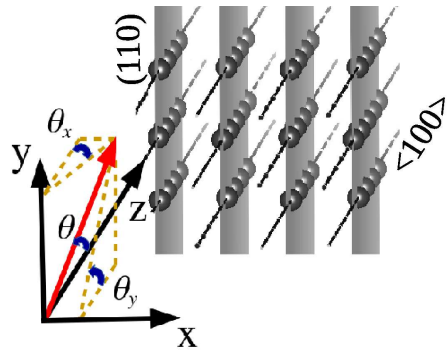}
\caption{\label{Fig1} Coordinate system to describe particle incidence on both the $\langle$100$\rangle$ atomic string and the (110) plane in a Si crystal.}
\end{figure}

\begin{equation}
\frac{d \sigma}{d\Omega}=|\sum_{j=1}^N e^{i \mathbf{qr_j}}|^2 \frac{d \sigma_{at}}{d\Omega},
\label{Eq1}
\end{equation}
where $\mathbf{q}=\mathbf{p} - \mathbf{p'}$ is the momentum
transferred to the crystal lattice, $N$ is the number of
scattering centers in a crystal and $d \sigma_{at}$ the
cross-section of scattering on a single atom. Hereinafter we will use the Coulomb screened (Yukawa) atomic potential \cite{Mol,GEANT}, to calculate this cross-section. By assuming that
atoms vibrate independently across their equilibrium positions,
the total differential cross section (\ref{Eq1}) can be split
into two parts \cite{TM,Landau}

\begin{equation}
\frac{d \sigma}{d\Omega}=\frac{d \sigma_{coh}}{d\Omega}+\frac{d \sigma_{inc}}{d\Omega},
\label{Eq2}
\end{equation}
where

\begin{equation}
\frac{d \sigma_{coh}}{d\Omega}=D |\sum_{j=1}^N e^{i \mathbf{qr_{j0}}}|^2 \frac{d \sigma_{at}}{d\Omega}
\label{Eq3}
\end{equation}
is the "coherent" term of the differential cross section
\cite{TM,Landau} with the same form as the
Laue-Bragg formula. This term represents the scattering process
without any energy transfer to the crystal lattice, the
probability of which is given by the Debye-Waller factor
$D=exp(-q^2 u_1^2)$, where $u_1$ is the one-dimensional amplitude of thermal oscillations $u_1^2=<\mathbf{u_j^2}>/3$. In turn, the "incoherent" term of the
scattering cross section

\begin{equation}
\frac{d \sigma_{inc}}{d\Omega}=N(1-D) \frac{d \sigma_{at}}{d\Omega} = \frac{d \sigma_{am}}{d\Omega} - \frac{d \sigma_{1}}{d\Omega}
\label{Eq4}
\end{equation}
is proportional to the probability $1-D$, that an energy
transfer occurs, and can be naturally separated into two
contributions

\begin{equation}
\frac{d \sigma_{am}}{d\Omega}=N \frac{d \sigma_{at}}{d\Omega} \text{ and } \frac{d \sigma_{1}}{d\Omega}=N D \frac{d \sigma_{at}}{d\Omega},
\label{Eq5}
\end{equation}
where $d \sigma_{am}/d \Omega$ represents the scattering in an amorphous media, $d \sigma_1/d \Omega$ the difference between amorphous and crystal cases in differential cross-section of incoherent scattering unaccompanied by the atom vibration state change.

As the atom positions $\mathbf{r_{j0}}$ are uncorrelated as in an
amorphous medium, the cross section (\ref{Eq3}) exactly compensates for $d \sigma_1/d \Omega$
(\ref{Eq5}), making the total particle scattering cross section
(\ref{Eq2}) both equal to $d \sigma_{am}/d \Omega$ (\ref{Eq5}) and
independent of $D$. However, $d \sigma_{coh}/ d\Omega$ (\ref{Eq3}) does not necessarily
compensate for $d \sigma_1/d \Omega$ (\ref{Eq5}) in crystals. Indeed, at certain particle incidence direction, $d \sigma_{coh}/ d\Omega$ can be practically
nullified, manifesting thus the \textit{coherent scattering
suppression} (\textit{CSS}) effect:
\begin{equation}
d \sigma \xrightarrow{d \sigma_{coh}=0} d \sigma_{am}-d \sigma_{1}. 
\label{Eq6}
\end{equation}

In the case of a Si crystal oriented as in Fig. \ref{Fig1}, the conditions in Eq. (\ref{Eq6}) for observation of the CSS effect, are naturally about 
the angles of incidence w.r.t. $\langle
100\rangle$ axis, i.e. $\theta=\sqrt{\theta_x^2+\theta_y^2} \approx
\theta_y$ and (110) Si plane, i.e. $\theta_x$. They should be chosen in the regions, respectively
\begin{equation}
\frac{10 \theta_{ch}^{pl} d_{ax}}{\pi u_1} \approx 18\text{
}mrad  < \theta_y < \frac{\pi u_1 }{d_{at}} \approx
43\text{ }mrad, \label{Eq13}
\end{equation}
where $d_{at}$ is the interatomic distance in the
$\langle 100\rangle$ axis, and
\begin{equation}
10 \theta_{ch}^{pl} \approx 2\text{
}mrad  < \theta_x < \frac{\pi u_1 }{d_{ax}}\theta_y \approx 4.3\text{ }mrad,
\label{Eq14}
\end{equation}
where $d_{ax}$ is the interaxial distance for the (110) plane, $\theta_{ch}^{pl}$ for planar channeling critical angle. The left hand sides of both Eqs. (\ref{Eq13}) and (\ref{Eq14}) assure the necessary precision of straight line approximation \cite{Bazylev}. Maximal CSS effect is attained at $\theta_y = 34.9$ $mrad$ ($2^\circ$) w.r.t. $\langle$100$\rangle$ Si axis and $\theta_x = 3$ $mrad$ w.r.t. (110) Si plane. More information about this choice as well as about Born approximation theory on CSS effect is consider in appendix A. 

\section{Experimental results}

\begin{figure}
\includegraphics[width=0.48\textwidth]{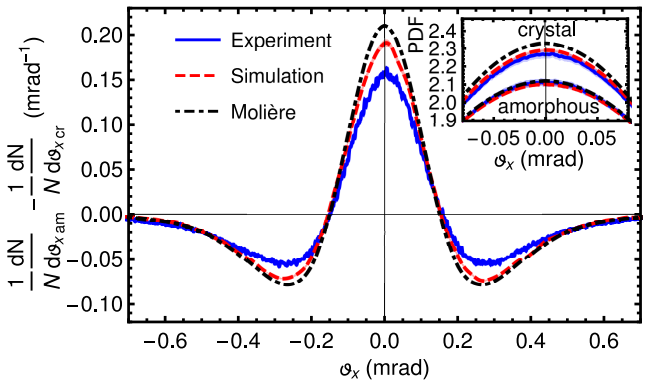}
\caption{\label{Fig3} The difference of the angular distributions of deflected beam
by a silicon crystal, aligned at $\theta_x = 3$ $mrad$, $\theta_y
= 34.9$ $mrad$ ($2^\circ$) and amorphous
membranes for experiment (solid), simulations (dashed) and Moli\`{e}re model
(dot dashed). The distributions themselves are placed in insert. The curve width represents the errors.}
\end{figure}

Direct observation of CSS effect was carried out at the MAinz MIkrotron (MAMI) by scattering of 855 MeV electrons on either crystal or amorphous Si plate $l \sim 30$ $\mu m$ thick along the beam. The experimental setup is described in \cite{MAMI}. The crystalline plate was fabricated via anisotropic wet etching procedures \cite{mirroring}, while the amorphous plate was produced by sputtering deposition of silicon onto a dedicated substrate, followed by anisotropic wet-chemical erosion to arrive at a thin amorphous plate surrounded by a bulky Si frame for mechanical stability \cite{sputter}. Thickness was $34.2 \pm 0.2$ $\mu m$ for the crystal plate and $32.76 \pm 0.09$ $\mu m$ for the amorphous one. Following the method \cite{Creagh} with a high resolution x-ray diffractometer, we verified their structure and measured that the two plates differ in \textit{mass} thickness by 0.4 \% only. The angular distributions of horizontal scattering angles $\vartheta_x$ for both kinds of plates as well as the difference of these distributions are illustrated in Fig. \ref{Fig3}.
  
\begin{table}[b]
\centering
\caption{\label{tab:table}%
Moli\`{e}re theory parameters for silicon crystal ($\theta_x=3$ $mrad$, $\theta_y = 34.9$ $mrad$ ($2^\circ$)) and amorphous
plates, calculated by (\ref{Eq12}-\ref{Eq10}) and by fitting
both experimental and simulation data}
\begin{tabular*}{\columnwidth}{@{\extracolsep{\fill}}llll@{}}
\hline
\textrm{$\vartheta$ ($\mu rad$)}&
\textrm{experiment}&
\textrm{simulation}&
\textrm{Moli\`{e}re theory}\\
\hline
$\vartheta_s^{cr}$& $183.4\pm 2.0$ & $181.5\pm 0.6$ & $178.6\pm 0.6$\\
$\vartheta_s^{am}$& $196.5\pm 2.0$ & $197.3\pm 0.3$ & $195.6\pm 0.3$\\
\textrm{$\vartheta_{min}^{cr}$}& $ 17.3 \pm 0.8 $ &  $ 18.0 \pm 0.4 $ &  $ 19.25 $\\
\textrm{$\vartheta_{min}^{am}$}& $ 12.5 \pm 0.7 $ & $ 12.30 \pm 0.13 $ & $ 12.83 $\\
\hline
\end{tabular*}
\end{table}

\begin{figure*}
\includegraphics[width=0.99\textwidth]{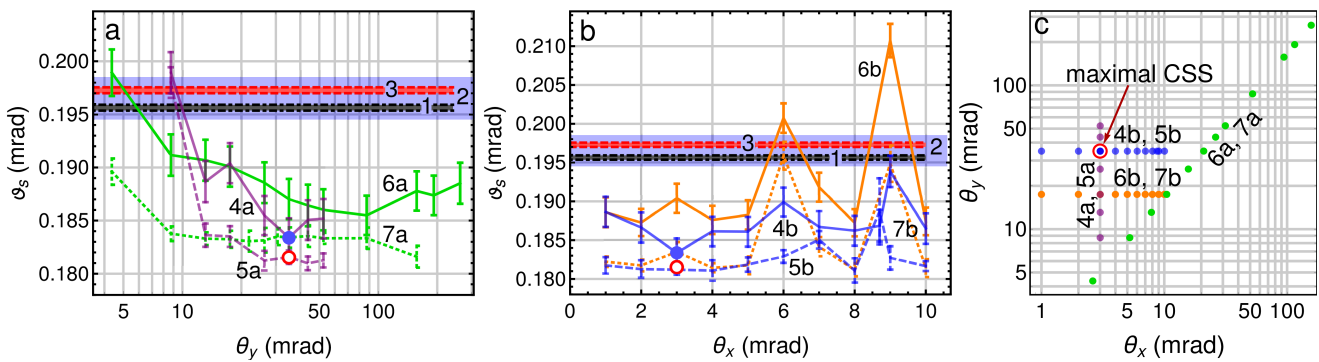}
\caption{\label{Fig4} Dependence of angle $\vartheta_s$ on the vertical (a) and horizontal (b) angles of crystal orientation. The lines 1--3 are the results for the amorphous target by the Moli\`{e}re model, experiment and simulations, respectively. Curve pairs (4,5), (6,7) represent the results for the crystal target. Even numbers are for experiment (solid curves), odd numbers are for CRYSTAL simulations (dashed curves).
a: (4a, 5a) $\theta_x$ is fixed at $3$ $mrad$; (6a, 7a) both $\theta_x$ and $\theta_y$ are varied while keeping the ratio  $\theta_x/\theta_y =0.6$ constant to avoid the intersection with high miller index crystalline planes. b: $\theta_y$ is fixed (4b, 5b) at $34.9$ $mrad$ ($2^\circ$); (6b, 7b) to $17.5$ $mrad$ ($1^\circ$). Blue disks (experiment) and red circles (CRYSTAL simulations) are the results for crystal alignment for maximal CSS effect $\theta_x = 3$ $mrad$, $\theta_y = 34.9$ $mrad$ highlighted in Fig. \ref{Fig3} and Table \ref{tab:table}. The errors are given by the thickness of the lines for the amorphous target as well as by the error bars for the crystal. c: the values of angles $\theta_x$ and $\theta_y$, used in corresponding curves in (a) and (b).}
\end{figure*}

We used Moli\`{e}re theory \cite{Mol,Bethe} to calculate the distribution of horizontal scattering angles $\vartheta_x$
\begin{equation} 
\begin{array}{l}
p(\vartheta_x)=\frac{1}{\pi} \int\limits_{-\infty}^{\infty}\! \sum_{j=0}^{\infty} \left( \frac{\vartheta_s}{\vartheta_c} \right)^{2j} f^{(j)} \left(\sqrt{\frac{\vartheta_x^2+\vartheta_y^2}{2 \vartheta_s^2}}\right) d \vartheta_y,\\
\vartheta_c = \alpha [Z(Z+1) z 4 \pi n_{at} l]^{1/2}/pv,
\end{array}
\label{Eq12}
\end{equation}
where $p$ and $v$ are the particle momentum and velocity respectively, $\alpha$ the fine structure constant, $n_{at}$ the atomic density, $Z$ the atomic number, $z$ the particle charge in $e$ units, the functions $f^{(j)}$ are introduced in \cite{Mol,Bethe}.

$\vartheta_s$ is the only free parameter of Moli\`{e}re theory, which is of the order of the r.m.s. of distribution $p(\vartheta_x)$. Indeed, at lowest order ($j=0$), $\vartheta_s$ is exactly the r.m.s. of the distribution. For comparison with the experimental results, $p(\vartheta_x)$ was convoluted with the angular distribution of the beam characterized by $27$ $\mu rad$ divergence. In Table \ref{tab:table} $\vartheta_s$ is reported for 3 different cases: by fitting of experimental data with Eq. (\ref{Eq12}), by fitting in the same way of the distributions obtained by Monte Carlo simulations described below and by a full theoretical estimation also described below. The difference between $\vartheta_s$ for the crystalline and amorphous cases are reported for all these 3 cases in Fig. \ref{Fig3} and showing a firm observation of the CSS.

According to Moli\`{e}re theory \cite{Mol,Bethe}, the theoretical value $\vartheta_s$ is defined as a function of an effective minimal scattering angle $\vartheta_{min}^{am} = \hbar /Rp$, evaluated through the Coulomb screened (Yukawa) atomic potential, where $R = a_{TF}(1.13+3.76(\alpha Z z /\beta )^2)^{-1/2}$ is the atom screening radius, chosen according to \cite{Mol,Bethe}, $a_{TF}=0.8853 a_B Z^{-1/3}$ the Thomas-Fermi screening radius, $a_B$ the Bohr radius. The CSS effect in crystals can be similarly described by introducing the effective minimal scattering angle $\vartheta_{min}^{cr}$, obeying the relation  

\small
\begin{equation}
\ln \vartheta_{min}^{cr} = \ln \vartheta_{min}^{am} +
\frac{0.5Z}{Z+1} \left[\left(1+\frac{u_1^2}{R^2}\right)
e^{\frac{u_1^2}{R^2}} E_1 \left(\frac{u_1^2}{R^2}\right)-1
\right], \label{Eq10}
\end{equation}
\normalsize 
obtained by integration of the cross section (\ref{Eq4}) following the Moli\`{e}re theory \cite{Mol,Bethe} at the condition of complete coherent scattering suppression $d \sigma_{coh}/ d\Omega=0$. $E_1$ is the exponential integral. The last item in (\ref{Eq10}) is representation of the $d \sigma_1/d \Omega$ contribution in (\ref{Eq4}), manifesting the difference in incoherent scattering between the crystalline and amorphous cases. The ratio $Z/(Z+1)$ is necessary to exclude from the CSS effect scattering on atomic electrons, described by ``+1'' in the definition of $\vartheta_c$ (see Eq. (\ref{Eq12})) \cite{Bethe}.

According to Eq. (\ref{Eq10}), the CSS effect can be interpreted as the increase in the effective minimal scattering angle from $\vartheta_{min}^{am}$ to $\vartheta_{min}^{cr}$. Both of these values as well as the corresponding angles $\vartheta_s$ are summarized in Table \ref{tab:table} and used for calculation of the theoretical angular distributions (Moli\`{e}re theory) by Eq. (\ref{Eq12}) as shown in Fig. \ref{Fig3}. The experimental and simulated values for $\vartheta_{min}^{am}$ and $\vartheta_{min}^{cr}$ shown in Table \ref{tab:table} were calculated from the corresponding values $\vartheta_s$ in Table \ref{tab:table}.

Both the experiment and the theory demonstrate the extent of CSS effect although the agreement is not perfect. However, neither Born approximation nor Moli\`{e}re theory, modified using Eq. (\ref{Eq10}), take into consideration the residual coherent effects, such as the influence of the string potential on incoherent scattering and an additional angular spread arising at  both the entrance and exit of a particle in/out of a crystal. All these effects have been fully taken into consideration by using the CRYSTAL code \cite{Syt1,Syt2}, a program for Monte Carlo simulations of charged particle trajectories in the potential of a crystalline medium with simulation of incoherent scattering, tested in different experiments \cite{Mazz,Ban,Guidi}. The results of simulations are shown in Fig. \ref{Fig3} and Table \ref{tab:table}.

The CSS effect was also investigated both experimentally and by simulations over a wide range of particle incidence directions $\theta_y$ and $\theta_x$ including the values beyond the optimal angular range of Eqs. (\ref{Eq13}--\ref{Eq14}). 

The corresponding angular distributions were used for a fit of Moli\`{e}re theory to calculate dependence of $\vartheta_s$ on both $\theta_y$ and $\theta_x$, as shown in Figs. \ref{Fig4}a and \ref{Fig4}b, respectively. The horizontal lines indicate the $\vartheta_s$ values evaluated for the amorphous case using Moli\`{e}re theory. The values of $\theta_y$ and $\theta_x$ used in Fig. \ref{Fig4}a and \ref{Fig4}b are shown in Fig. \ref{Fig4}c.

Fig. \ref{Fig4}a demonstrates that although the full CSS effect is attained within a relatively narrow angular region as in Eqs. (\ref{Eq13}--\ref{Eq14}), a considerable partial CSS is still observed at least up to $\theta_x \sim 157$ $mrad$ ($9^\circ$), $\theta_y \sim 262$ $mrad$ ($ 15^\circ$) (see Fig. \ref{Fig4}a). Since the other axes can contribute at such high angular values, some partial CSS effect can appear at any crystal alignment.

Fig. \ref{Fig4}b reveals that the CSS effect may be hampered at relatively large incidence angles by the appearance of the peaks originating from the coherent scattering on high-index planes, which were observed experimentally as well as predicted by simulations especially in scan 6b at 6 and 9 mrad of $\theta_x$.

Thereby, we conclude that the anisotropy of multiple scattering can be observed at broad angle w.r.t. the crystalline orientations. Furthermore, at higher energies the CSS effect is preserved, since neither the right hand side of Eq. (\ref{Eq10}) nor the upper angular limits in (\ref{Eq13}-\ref{Eq14}) depend on energy for ultrarelativistic particles, while the lower limits decrease further with energy. By the same reason the anisotropy of multiple scattering of ultrarelativistic particles depends neither on the particle mass nor on its charge sign. However, the effect is attenuated as the particle charge rises, becoming negligible for high-$z$ ions.

\section{Conclusions}

In conclusion, the maximal effect of CSS in crystals has been observed under proper choice of the incidence angle. It resulted in decrease by about 7 \%
in the multiple scattering angle in a thin Si crystal target with the same mass thickness as an amorphous one.

Partial CSS does exist and was observed even up to about $15^\circ$ tilt from the crystal axis, owing to its independence on energy, mass and charge sign. We believe that the angular dependence of multiple scattering may be useful knowledge to aid the modeling and design of nuclear physics and high-energy experiments. For instance, Geant4 \cite{GEANT}, currently using the Moli\`{e}re model of scattering \cite{Mol,Bethe}, may be implemented with the model developped in this work.

\begin{acknowledgements}
We acknowledge partial support of the INFN-ELIOT experiment and by the European Commission (the PEARL Project within the H2020-MSCA-RISE-2015 call, GA 690991). E. Bagli, L. Bandiera and A. Mazzolari recognize the partial support of FP7-IDEAS-ERC CRYSBEAM project GA n. 615089. A. Mazzolari and V. Guidi acknowledge founding from PRIN 2015LYYXA8 ``Multi-scale mechanical models for the design and optimization of micro-structured smart materials and metamaterials''. We also acknowledge the CINECA award under the ISCRA initiative for the availability of high performance computing resources and support. We acknowledge Professor H. Backe for fruitful discussions.
\end{acknowledgements}

\appendix

\section{Theory of multiple scattering in a crystal}

In this appendix we apply Born approximation \cite{Landau,TM,Baier} to establish the conditions, in particular for crystal alignment, for coherent scattering suppression in a crystal owing to by the discreteness of momentum transfer, in particular Eqs. (7) and (8) in the main text. Calculation was done in the case of $<$100$>$ axis and (110) plane of a Si crystal. 

In order to find the conditions for coherent scattering suppression, one should compare the r.m.s. angle $\langle \vartheta_{x'}^2 \rangle_{coh}$\footnotemark \footnotetext{$x'y'z'$ is the Cartesian coordinate system connected with an incident particle, the momentum $\overrightarrow{p}$ of which is parallel to $z'$}  of coherent scattering with the r.m.s. angle $\langle \vartheta_1^2 \rangle$, characterizing the suppression of incoherent scattering, established by the cross-sections defined in the paper as $d\sigma_{coh}$ and $d\sigma_1$, respectively. In other words, the difference of the total r.m.s. scattering angle in amorphous matter and in a crystal is $\langle \vartheta_{x'}^2 \rangle_{coh}-\langle \vartheta_1^2\rangle$. The condition for complete coherent scattering suppression holds 
\begin{equation}
\label{A1}
\langle \vartheta_{x'}^2 \rangle_{coh}/\langle \vartheta_1^2\rangle \ll 1 .
\end{equation}
$\langle \vartheta_1^2 \rangle$ is obtained by integration of the cross-section $d\sigma_1$ as follows     

\begin{eqnarray}
\label{A2}
\left\langle \vartheta_{x'}^2\right\rangle_1 =  n_{at} l \int \vartheta _{x'}^2 \frac{d\sigma_1}{d\Omega} \, d\Omega =n_{at} l \int \frac{q_{x'}^2}{p^2} \frac{d\sigma_1}{d\Omega} \, d\Omega = \nonumber \\ 
\frac{\theta_1^2}{2} \left[\left(1+\frac{u_1^2}{R^2}\right)
e^{\frac{u_1^2}{R^2}} E_1 \left(\frac{u_1^2}{R^2}\right)-1\right],
\end{eqnarray} 
see definitions of such quantities in the paper.

Similarly, one can calculate the angle $\langle \vartheta_{x',y'}^2 \rangle_{coh}$ by using the cross-section $d\sigma_{coh}$. With this aim, one can express a direct summation of the exponents over the $8 N^3$ atoms of $N^3$ elementary cells as

\begin{eqnarray}
\label{A3}
\left|\sum _{j=1}^{8 N^3} e^{\text{iqr}_{\text{i0}}}\right|^2  = \left(\frac{2 \pi  N}{d_{at}}\right)^3 \sum _{n_1=-\infty}^{\infty} \sum _{n_2=-\infty}^{\infty} \sum _{n_3=-\infty}^{\infty} |S|^2 \times \nonumber \\ \delta \left(q_1-\frac{2 \pi  n_1}{d}\right) \delta \left(q_2-\frac{2 \pi  n_2}{d}\right) \delta \left(q_3-\frac{2 \pi  n_3}{d}\right) ,
\end{eqnarray} 
where $d_{at}$ is lattice spacing and $S$ the structure factor, which can be written for diamond-type crystal lattice [9, 18] as
\begin{eqnarray}
\label{A4}
S=\left[1+e^{i \pi \left(n_1+n_2+n_3\right)/2}\right] [1+\cos \left(\pi  \left(n_1+n_2\right)\right)+ \nonumber \\ \cos \left(\pi  \left(n_2+n_2\right)\right)+\cos \left(\pi  \left(n_1+n_3\right)\right) ].
\end{eqnarray}
1, 2 and 3 correspond to the Cartesian axes associated with the canonical base. In addition, one should [6] average the cross section $d\sigma_{coh}/d\Omega$ over initial beam angular distribution. For certain cylindrical Gaussian, the following angular distribution will be used
\begin{equation}
\label{A5}
\frac{1}{N} \frac{d N}{d \overrightarrow{\theta }}=\frac{1}{2 \pi  \delta ^2} e^{-\left(\overrightarrow{\theta }-\overrightarrow{\theta_0 }\right)^2/2\delta ^2},
\end{equation}
where $\overrightarrow{\theta}= (\overrightarrow{\theta_1}, \overrightarrow{\theta_2})$ are the angles of integration,  $\overrightarrow{\theta_0}= (\overrightarrow{\theta_{01}}, \overrightarrow{\theta_{02}})$ the mean angles of beam direction w.r.t. the crystal planes 23 and 13 respectively. Thereby, $\theta_0$ is the beam angle w.r.t. the axis 3. Both angles are hereinafter considered as small angles: $\theta \ll 1$, $\theta_0 \ll 1$. $\delta$ is the r.m.s. angular divergence. The actual value of $\delta$ can vary in the range of $\sim20-200$ $\mu$rad through the passage of a $\sim 30$ $\mu$m Si target. Since the typical incidence angles well exceed 1 mrad $ \gg \delta$, the beam angular divergence can be treated as negligible and assumed to be independent on particle penetration depth. 

By substituting $d\sigma_{coh}$ for $d\sigma_1$ in (\ref{A2}) and by using (\ref{A3}), (\ref{A4}), (\ref{A5}) one finally obtains
\begin{eqnarray}
\label{A6}
\begin{array}{l}
\left \{
\begin{array}{c}
 \langle \vartheta_{x'}^2 \rangle_{coh} \\
 \langle \vartheta_{y'}^2 \rangle_{coh} \\
\end{array}
\right \} = \int \left \{
\begin{array}{c}
\vartheta_{x'}^2 \\
\vartheta_{y'}^2 \\
\end{array}
\right \} \frac{d\sigma_{coh}}{d\Omega} \, d\Omega = \frac{1}{p^2} \int \left \{
\begin{array}{c}
q_{x'}^2 \\
q_{y'}^2 \\
\end{array}
\right \} \frac{d\sigma_{coh}}{d\Omega} \, d\Omega = \\[.4cm]
\left(\frac{2 \alpha  Z}{\theta _0 p v}\right)^2 \left(\frac{2 \pi}{d_{at}}\right)^3 \frac{n_{at}}{8} l  \sum _{n_1=-\infty}^{\infty} \sum _{n_2=-\infty}^{\infty} \sum _{n_3=-\infty}^{\infty} |S|^2 \times \\[.4cm] \left \{
\begin{array}{c}
(\theta_{01} q_{2n_2} - \theta_{02} q_{1n_1})^2 \\
q_{3n_3}^2 \\
\end{array}
\right \} \frac{q_{n_1 n_2}}{q_{n_1 n_2 n_3}} \frac{\exp \left(-q_{n_1 n_2 n_3}^2 u_1^2 \right)}{\left(q_{n_1 n_2 n_3}^2+R^{-2} \right)^2} \times \\[.4cm] \frac{1}{\sqrt{2 \pi} \delta q_{n_1 n_2 n_3}} \exp \left[-(q_{3n_3} + q_{1n_1} \theta_{01} + q_{2n_2} \theta_{02})^2/2 \delta^2 q_{n_1 n_2}^2\right],
\end{array}
\end{eqnarray}
where $q_{n_1 n_2}^2=q_{1n_1}^2+q_{2n_2}^2$ and $q_{n_1 n_2 n_3}^2=q_{n_1 n_2}^2+q_{3n_3}^2$. It is important to underline, that Eq. (\ref{A6}) is valued for any type of a crystal lattice as well as for any axes. Then, the structure factor is set as well as the directions of 1,2,3 are assigned for the particular case under investigation. To simplify the integration of (\ref{A6}) on angles $d\theta_1$ $d\theta_2$, these angles were transformed by rotation of the coordinate system around axis 3 by angle $\varphi$, for which $\cos \varphi = q_{1n_1}/ q_{n_1 n_2}$, $\sin \varphi = q_{2n_2}/q_{n_1 n_2}$, and in which the second component of $q$ is equal to 0. We also applied the momentum conservation law $q_{1n_1} \theta_{01} + q_{2n_2} \theta_{02} = q_{n_1 n_2} \theta \cos \varphi = -q_{3n_3} \cos \theta$ ($p'^2 = (\overrightarrow{p} + \overrightarrow{q})^2 = p^2$).

\begin{figure*}
\includegraphics[width=0.33\textwidth]{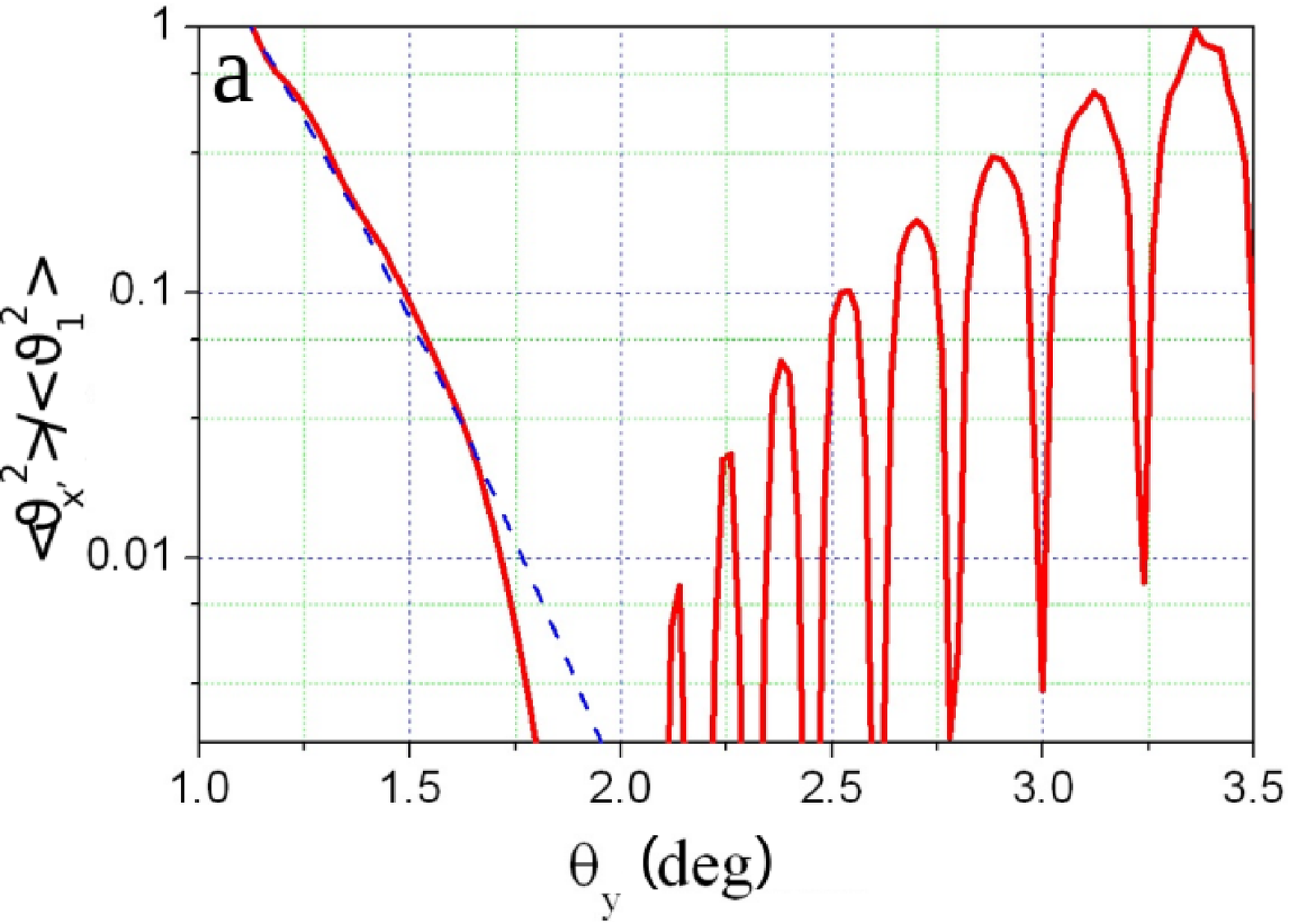}
\includegraphics[width=0.33\textwidth]{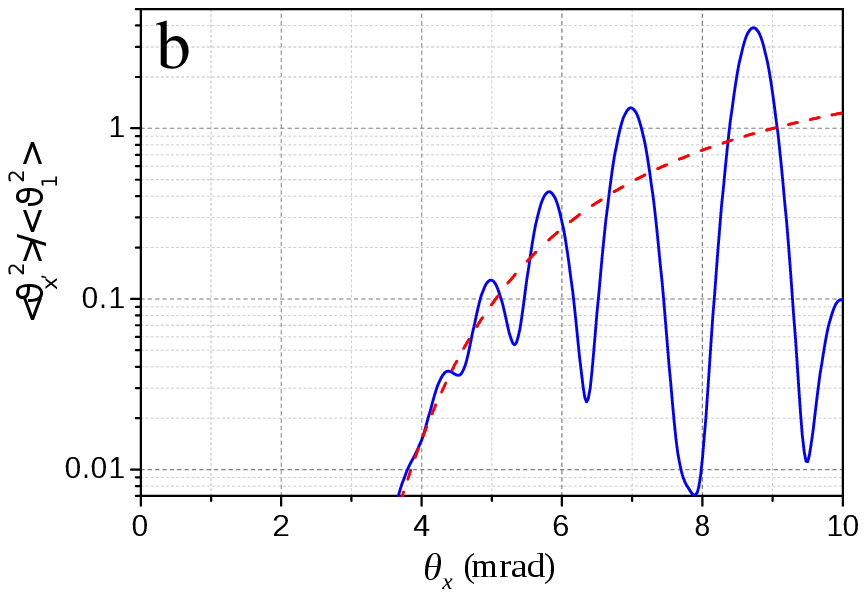}
\includegraphics[width=0.33\textwidth]{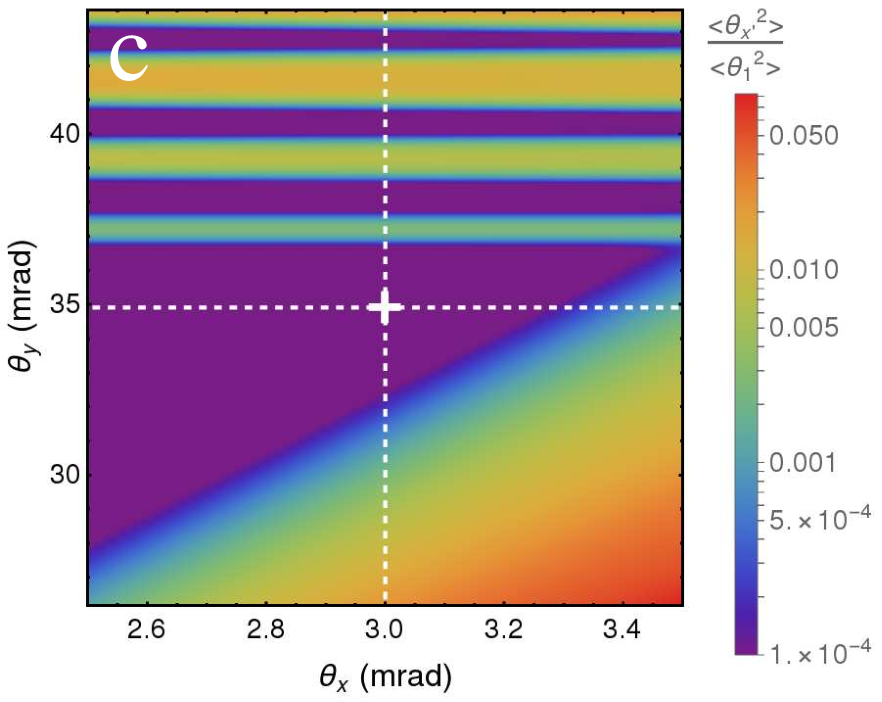}
\caption{\label{Fig1A} ``Planar'' and ``axial'' dependence of the average squared angle $\langle \vartheta_{x'}^2 \rangle_{coh}/\langle \vartheta_1^2\rangle $ (calculated with Eqs. (\ref{A6}) and (\ref{A2})) of coherent 855 MeV electrons scattering by (110) planes in a 30 $\mu m$ Si crystal on both incidence angles, $\theta_x$ and $\theta_y$ respectively. 
(a): ``Axial'' dependence, evaluated at $\theta_x = 3$ mrad with both $q_z = 0$ and all the $q_z \neq 0$ contributions (red solid line), the left dashed blue line is the $q_z = 0$ contribution (\ref{A10}).
(b) ``Planar'' dependence, evaluated at $\theta_y = 34.9$ mrad ($2^\circ$) with both $q_z = 0$ and all the $q_z \neq 0$ contributions (blue solid line), while the red dashed line was evaluated in the limit of isolated plane (\ref{A13}).
(c): The same dependence on both $\theta_x$ and $\theta_y$ angles. White cross marks the optimal orientation for coherent scattering suppression $\theta_x = 3$ mrad and $\theta_y = 34.9$ mrad, dashed horizontal and vertical lines represent solid curves on plots (a) and (b) respectively.}
\label{FigA1}
 \end{figure*}

Here $\varphi$ is the angle formed by the transferred momentum projection onto the plane 12 $\overrightarrow{q}_{n_1 n_2}$ and the vector $\overrightarrow{\theta}$. One can pass to the same law in the negligible divergence limit $\delta \rightarrow 0$, arriving to the condition 
\begin{equation}
\label{A7}
q_{n_1 n_2 n_3}^2 = q_{3n_3}^2 \left(1+\cos ^2\varphi/\theta^2\right) > q_{3n_3}^2/\theta^2,
\end{equation}
which limits the momentum transfer value from below at $q_{n_3} \neq 0$. By rewriting Eq. (\ref{A7}) in terms of the Debye-Waller factor, which presents in (\ref{A6})
\begin{equation}
\label{A8}
D = \exp(-u_1^2 q^2) \leq \exp(-u_1^2 q_{3n_3}^2/\theta^2),
\end{equation}
we conclude that the discrete nature of the longitudinal momentum transfer $q_{n_3} = 2 \pi n_3/d_{at}$ results in coherent scattering suppression by the Debye-Waller factor at $q>\hbar/u_1$. By assuming that $D \leq 0.01$, and substituting the minimal momentum axial transfer value $q_{31}=2 \pi /d_{at}$, one readily comes to the first condition for coherent scattering suppression
\begin{equation}
\label{A9}
\theta<\frac{u_1 q_{31}}{2} = \frac{\pi u_1}{d_{at}} \approx 43\text{ }mrad\text{ }(2.5^\circ).
\end{equation}
This condition represents \textit{coherent scattering suppression by individual atoms in the strings}, when separate atomic strings scatter charged particles as a whole. The suppression range (\ref{A9}) is illustrated in Fig. \ref{FigA1}a, representing the ratio $\langle \vartheta_{x'}^2 \rangle_{coh}/\langle \vartheta_1^2\rangle$.

If the condition (\ref{A9}) is fulfilled, one can neglect the component $q_{3n3}$ in (\ref{A6}). By assigning the coordinate system 1,2,3 to the system $x,y,z$, in which $z$ is parallel to $\langle$100$\rangle$ Si axes as well as the planes (-110) and (110) are formed by xz and yz planes, respectively, and by using the structure factor (4A), one can rewrite (6A) as
\begin{eqnarray}
\label{A10}
\begin{array}{l}
\langle \vartheta_{x'}^2 \rangle_{coh} = \int \vartheta_{x'}^2 \frac{d\sigma_{coh}}{d\Omega} \, d\Omega = \frac{1}{p^2} \int q_{x'}^2 \frac{d\sigma_{coh}}{d\Omega} \, d\Omega = \\[.4cm]
\left(\frac{2 \alpha  Z}{\theta _0 p v}\right)^2 \frac{2 \pi}{d_{at}}\frac{2 \pi}{d_{ax}}\frac{2 \pi}{d_{pl}} n_{at} l \times \\[.4cm]  \sum _{m=-\infty}^{\infty} \sum _{k=-\infty}^{\infty} (\theta_{0x} q_{ym} - \theta_{0y} q_{xk})^2 \frac{\exp \left(-q_{mk}^2 u_1^2 \right)}{\left(q_{mk}^2+R^{-2} \right)^2} \times \\[.4cm] \frac{1}{\sqrt{2 \pi} \delta q_{mk}} \exp \left[-(q_{xk} \theta_{0x} + q_{ym} \theta_{0y})^2/2 \delta^2 q_{mk}^2\right],
\end{array}
\end{eqnarray}
where $q_{mk}^2 = q_{xk}^2 + q_{ym}^2$, $d_{pl} = d_{ax} = d' = d_{at}/2\sqrt{2}$, $q_{xk}=2 \pi k/ d_{pl}$ and $q_{ym}=2 \pi m/d_{ax}$, $k, m = 0, \pm 1, \pm 2, ..$.
Here the structure factor is trivial, i.e. $|S|^2 = 1$. Eq. (\ref{A10}) represents the case $q_z = 0$, defined by the condition (\ref{A9}). This equation was used for calculation of the dependence of the average squared angle $\langle \vartheta_{x'}^2 \rangle_{coh}/\langle \vartheta_1^2\rangle $ on $\theta_y$ in the limit $q_z = 0$.

The momentum consevation law can be obtained similarly by Eq. (\ref{A7}) under the limit of negligible divergence $\delta \rightarrow 0$, $q_{xn} = -q_{ym} \theta_y/\theta_x$, $q_{mk}^2 = q_{xk}^2 + q_{ym}^2=q_{ym}^2 \theta^2/\theta_x^2$, giving the condition for the Debye-Waller factor (compare (\ref{A8})):
\begin{eqnarray}
\label{A11}
D = \exp(-u_1^2 q^2) = \exp \left[-u_1^2 (q_x^2+q_y^2) \right] \leq \nonumber \\ \exp(-u_1^2 q_{y1}^2 \theta^2/\theta_x^2).
\end{eqnarray}
Assuming again $D \leq 0.01$ one arrives to the second condition for coherent scattering suppression:
\begin{eqnarray}
\label{A12}
10 \theta_{ch}^{pl} \approx 2\text{
}mrad  < \theta_x < u_1 q_{y1} \theta_y/2 = \nonumber \\ \frac{\pi u_1 }{d_{ax}}\theta_y \approx 0.12 \theta_y \approx 4.3 mrad.
\end{eqnarray}
This condition represents \textit{coherent scattering suppression by individual strings in the planes}, when separate planes scatter charged particles as a whole. $q_{y1}=2 \pi / d_{ax} = 4 \sqrt{2} \pi /d_{at}$ is the minimal momentum transfer parallel to the yz or (110) plane. The numerical estimate (\ref{A12}) has been done for the chosen axial angle $\theta = 2^\circ$. The suppression range (\ref{A12}) is illustrated in Fig. \ref{FigA1}b, representing the ratio $\langle \vartheta_{x'}^2 \rangle_{coh}/\langle \vartheta_1^2\rangle$ in dependence on the horizontal angle $\theta_x$. The peaks in this dependence originate from the coherent scattering on high-index planes. On can eliminate their influence by the application of the limit isolated plane ($d_{pl} \rightarrow \infty$). Following condition (\ref{A11}) one can rewrite Eq. (\ref{A10}) as
\begin{eqnarray}
\label{A13}
\langle \vartheta_{x'}^2 \rangle_{coh} = \left(\frac{2 \alpha  Z}{\theta _0 p v}\right)^2 \frac{2 \pi}{d_{at}}\frac{2 \pi}{d_{ax}} n_{at} l \frac{\theta_x^2+\theta_y^2}{\theta_x^3} \times \nonumber \\ \sum _{m=-\infty}^{\infty} q_{ym}^2 \frac{\exp \left(-u_1^2 q_{ym}^2 (\theta_x^2+\theta_y^2)/\theta_x^2 \right)}{\left(q_{ym}^2 (\theta_x^2+\theta_y^2)/\theta_x^2+R^{-2} \right)^2}.
\end{eqnarray}
This equation was used to build up the dependence illustrated in Fig. \ref{FigA1}b in the limit of isolated plane, eliminating the contribution by high-index planes, but preserving the range of coherent scattering suppression, defined by the condition (\ref{A12}).

The left hand side condition (\ref{A12}), in which $\theta_{ch}^{pl}$ is the planar channeling angle, has been implied to assure sufficient precision of Born approximation. It follows, directly from the estimate of the influence of particle deflection in the average planar potential on the angle of incoherent electron scattering by nuclei [19], which cannot be estimated in Born approximation. The condition (\ref{A12}) can be substituted in (\ref{A9})

\begin{eqnarray}
\label{A125}
\frac{10 \theta_{ch}^{pl} d_{ax}}{\pi u_1} \approx 18\text{
}mrad\text{ }(1^\circ) < \theta_y \leq \theta<\frac{u_1 q_{31}}{2} = \nonumber \\ \frac{\pi u_1}{d_{at}} \approx 43\text{ }mrad\text{ }(2.5^\circ).
\end{eqnarray}

Fig. \ref{FigA1}c illustrates the region of practically complete suppression $\langle \vartheta_{x'}^2 \rangle_{coh} < 0.01 \langle \vartheta_1^2\rangle$ (as required by (\ref{A1})) of the coherent scattering (central dark triangle) predicted by Eqs. (\ref{A12}) and (\ref{A125}). This coordinate point   $\theta_x=3$ mrad and $\theta_y=2^\circ \approx 35$ mrad, marked by the white cross, corresponds to the electron incidence direction, which has been selected as the most appropriate one for the conducted experiment on the first observation of the coherent scattering suppression effect, presented in Fig. 2 of the paper.



\end{document}